\documentclass{WileyMSP-template}

\usepackage{epstopdf}%This line makes .eps figures into .pdf - please comment out if not required.

\newcommand{\rezA}{\AA$^{-1}$}

\newcommand{\rezAs}{\AA$^{-1}$ \,}

\newcommand{\mums}{$\mu \mbox{m}$\,}
\newcommand{\mus}{$\mu \mbox{s}$}
\newcommand{\muss}{$\mu \mbox{s}$\,}
\newcommand{\Jm}{J/m$^2$}
\newcommand{\Jms}{J/m$^2$\,}

\begin{document}

\title{Pulsed-laser induced gold microparticle fragmentation by thermal strain$^{\dag}$}
\maketitle

% Author: Please give full first and last names for authors and include * after the name of all corresponding authors
\author{Yogesh Pokhrel}
\author{ Meike Tack}
\author{Sven Reichenberger*}
\author{Matteo Levantino}
\author{ Anton Plech*}

% Affiliations: Please provide adacemic titles (Prof. or Dr.) for all authors where applicable, and include an institutional email address for all corresponding authors
\begin{affiliations}
Y. Pokhrel, A. Plech\\
Institute for Photon Science and Synchrotron Radiation, Karlsruhe Institute of Technology, Hermann-von-Helmholtz-Platz 1, D-76344 Eggenstein-Leopoldshafen, Germany, EU\\
anton.plech@kit.edu \\

M. Tack, S. Reichenberger\\
Department of Technical Chemistry I and Center for Nanointegration Duisburg-Essen, University of Duisburg-Essen, Universit\"atsstrasse 7, D-45141 Essen, Germany, EU\\
sven.reichenberger@uni-due.de

M. Levantino \\
European Synchrotron Radiation Facility,  71, avenue des Martyrs, CS 40220, F-38043 Grenoble, France, EU
\end{affiliations}


%Please use \dag to cite the ESI in the main text of the article.
%If you article does not have ESI please remove the the \dag symbol from the title and the footnotetext below.
%\footnotetext{\dag~Supplementary Information available: [details of any supplementary information %available should be included here]. See DOI: 10.1039/cXCP00000x/}
%additional addresses can be cited as above using the lower-case letters, c, d, e... If all authors are from the same address, no letter is required

%%%END OF FOOTNOTES%%%

\begin{abstract}

Laser fragmentation of suspended microparticles is an upcoming alternative to laser ablation in liquid (LAL) that allows to streamline the the delivery process and optimize the irradiation conditions for best efficiency. Yet, the structural basis of this process is not well understood to date. Herein we employed ultrafast x-ray scattering upon picosecond laser excitation of a gold microparticle suspension in order to understand the thermal kinetics as well as structure evolution after fragmentation. The experiments are complemented by simulations according to the two-temperature model to verify the spatiotemporal temperature distribution. It is found that above a fluence threshold of 750 \Jms the microparticles are fragmented within a nanosecond into several large pieces where the driving force is the strain due to a strongly  inhomogenous heat distribution on the one hand and stress confinement due to the ultrafast heating compared to stress propagation on the other hand. The additional limited formation of small clusters is attributed to photothermal decomposition on the front side of the microparticles at the fluence of 2700 \Jm.

\end{abstract}

\section{Introduction}

Laser synthesis of nanoparticles in liquids has gained much attraction within the past years because of the prospect of producing a wide variety of particle materials not only in small on-demand batch processes but also in gram-scale synthesis under semi- or even fully continuous flow conditions \cite{streubel16,tack24b}. At the same time, several obstacles have to be overcome to define the product particles by size, chemical speciation, and yield of the process. The primary approach is to place a bulk target in a liquid and irradiate by short-pulse lasers, coined laser ablation in liquid (LAL). Size definition and speciation in LAL is limited due to the number of competing fundamental processes, including evaporation, spallation, thermomechanic disruption, bubble formation, or reactions of the solvent with limited ability to replace the materials between laser pulses. LAL typically produces multimodal size distributions, including size increase \cite{swiatkowska17} of the product colloid through repeated laser melting and fusing (LML) and redox reactions with the solvent \cite{amendola20,reich20} and by carbon shell formation in organic solvents \cite{amendola09}, respectively.

In order to harvest very small nanoparticles and clusters that show particularly promising catalytic or photonic activity, a second step of laser fragmentation in liquids (LFL) can be added to the process chain. LFL can be designed to operate on liquid jets that replace the colloid in a defined way and allow us to quantify the used laser fluence more easily \cite{zerebecki20} by masking than during LAL, which typically shows fluence gradients across the laser spot on the flat surface \cite{chen23} and ongoing erosion of the target \cite{letzel19} with variable efficiency \cite{reich19cpc} after continuous irradiation. Nanoparticle fragmentation has been understood very well recently due to combined efforts of large-scale simulations and experimental efforts to resolve the spatiotemporal evolution of colloid structure down to the atomic length scale \cite{ihm19,ziefuss20,plech24nano}. Full fragmentation of initial colloid particles of tens of nanometer size can be achieved by picosecond pulses by applying a laser fluence that exceeds 9-12 times the threshold fluence for lattice melting, still below the energy needed for full evaporation \cite{plech24nano}. The underlying mechanism is based on (thermal) phase explosion and spinodal decomposition. Single-particle studies have also revealed that a thermomechanic effect may play a role in the fragmentation of gold nanoparticles when the employed laser pulse duration (and hence the related, characteristic particle heating time) remains below the characteristic phonon propagation time (proportional to the speed of sound within the fragmented material). In such cases, the stress confinement criterium is fulfilled and fragmentation proceeds photomechanically, as observed through void formation in the particle center \cite{ihm19,Hwang2024}. Both, the experimental investigation and theoretical description of nanoparticle colloids is conceptually simplified because of the fact that energy deposition and flow within the particles is typically faster than structure formation, such as melting or decomposition. 

Recently, several groups have followed the approach of using commercially available microparticles as feedstock to produce the desired small nanoparticles \cite{wagener10,havelka21,Spellauge2023,Himeda2025,fromme25}. 
This approach has shown to robustly yield nanoparticles smaller than 3 nm, to allow for single-step laser processing but also to control the surface oxidation of the produced nanoparticles, when applying higher numbers of laser pulses. The  processing ability in a jet \cite{siebeneicher20} for a continuous and efficient material feed is important for defining the required mass-specific energy input and productivity, which is comparable to or even exceeds LAL \cite{tack24b,spellaugeunpublished}. 

While laser irradiation of microparticles in a liquid (LIML) may be viewed as a variant of LFL, in some aspects it shares processes that are found in LAL, such as inhomogeneous irradiation of the microparticle surface and tentative competition of processes pertinent to fragmentation as well as ablation.  

Paltauf and Schmidt-Kloiber \cite{paltauff99} have earlier set the scene for microparticle irradiation in liquid by optical stroboscopy. They describe liquid cavitation as well as stress confinement inside spheres of sub-millimeter sizes that leads to spallation by photo-acoustic stress from the center of the particles. Spherical particles build up a higher stress amplitude due to high symmetry as compared to cylindrically or irregularly formed particles. Kuzmin et al. \cite{kuzmin14} have described the fragmentation of microscale aluminum particles by a combined action of fission ("bisection") of the overexcited molten particles and thermal detachment of very small clusters. Sattari et al. \cite{sattari08} irradiated oxide microparticles in a gas stream and found an increased yield of nanoparticles below a size threshold of 100 nm with increased laser fluence. Spellauge et al. \cite{Spellauge2023} have recently combined visualization of single-microparticle fragmentation with quantification of product size distributions. The deduction of stress amplitudes through the observation of shock wave and cavitation bubble speed allowed us to conclude on the energy efficiency of the process and the transient state of the irradiated IrO$_2$ microparticles. In summary, irradiating microparticles with short laser pulses in the strain confinement regime is expected to produce a spatially inhomogeneous excitation profile across the microparticle as well as important strain waves and a more complicated fragmentation process than the one on nanoparticles with homogeneous excitation density \cite{plech24nano}. Elastic properties of the microparticles will play a more important role, such as brittleness \cite{tack24a}. 

We therefore performed picosecond-resolved pump-probe experiments on size-defined gold microparticles in a water jet by using pulsed x-rays as structural probe. The x-ray scattering allows us to resolve the thermal and melting dynamics of the particles with high time resolution as well as to watch the fragmentation process. Modeling of heat flow by numerical simulations according to the tow-temperature model of the electronic and phonon subsystems of the particles reveals the time scales for heat flow and particle melting. The experimental and theoretical fluences are well matched and show that fragmentation is caused by strain gradients across the particles. While fission into particles of tens to hundreds of nanometers prevails in the investigated low-fluence regime, fragmentation to nanometer-sized clusters plays a minor role and will set in at higher fluence. 

In summary, irradiating microparticles with short laser pulses in the strain confinement regime is expected to produce a spatially inhomogeneous excitation profile across the microparticle as well as important strain waves and a more complicated fragmentation process than the one on nanoparticles with homogeneous excitation density.

\section{Experimental details and analysis}

\begin{figure}[h]
\centering
  \includegraphics[width=12cm]{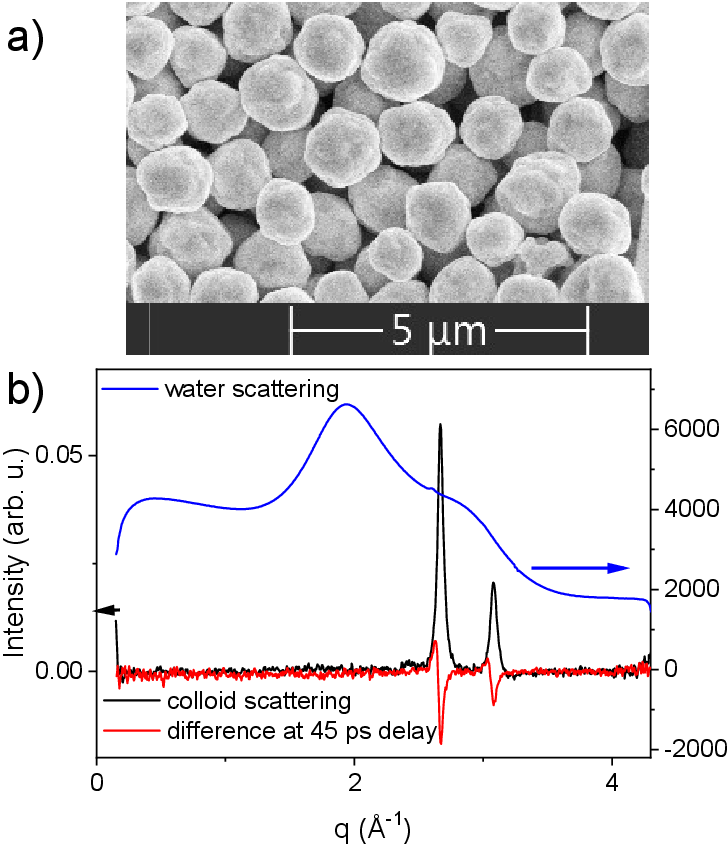}
  \caption{a) Scanning electron microscopy image of a powder of quasi-spherical gold microparticles. b) X-ray scattering distribution of an aqueous colloid containing 400 mg/l gold microparticles with full scattering including the water phase in blue, the extracted scattering from the gold particles in black, mainly showing the (111) and (200) gold powder peaks at 2.67 and 3.1 \rezAs, respectively. Upon laser excitation the powder peaks shift transiently and eventually are reduced in intensity as marked by the difference scattering at 45 ps delay between laser and x-ray pulses in red, respectively.}
  \label{SEM}
\end{figure}

{\bf Microparticle irradiation:} The gold microparticle powder with quasi-spherical particles and a narrow size distribution close to  1.2 \mums in diameter was purchased from Evochem and suspended in water by ultrasonic agitation. Scanning electron (SEM) images of the powder are displayed in fig. \ref{SEM} a. The particles settle within a ten minute time scale, but can easily be resuspended again by shaking. A  suspension (400 mg/l) was loaded into a polyethylene syringe that contained a magnetic stirrer bar to counter sedimentation. A syringe pump injects the suspension at a rate of 3.2 ml/min into an open-ended glass capillary (0.3 mm diameter, Hilgenberg). A free round jet is formed that is irradiated from the side by 1 ps laser pulses at 400 nm from a regenerative amplifier laser system (Coherent) at the beamline ID09 \cite{levantino21} at the European Synchrotron Radiation Facility (Grenoble, France). Coaxially to the direction of the focused laser beam (0.24 mm full width at half maximum), a pulsed x-ray beam probes at the same repetition rate of 1 kHz probes the structure of the sample. The laser fluence can be considered to be quite homogeneous within the probed area of the x-ray beam (0.04 mm) with an estimated 20 \% variation. The liquid was collected after irradiation at an incident fluence of 2700 \Jms and analyzed via SEM and transmission electron microscopy (TEM). For this analysis, the fragments are stabilized by 0.5 M NaOH at pH 10. Note that irradiated particles in the collected liquid are much less abundant than non-irradiated ones because of the difference in laser spot and jet size. Therefore, the size distributions and concentration are not representative of the abundance of fragmentation products but rather show the shape spectrum of fragments.   

{\bf Scattering signal:} The scattered x-rays are recorded on a 2D area detector (Rayonix HS170), whose distance from the sample can be varied to cover a scattering vector q$=$4$\cdot \pi/\lambda \cdot \sin (2\Theta/2)$ from 0.01 to 5 \rezAs, with $\lambda$ and $2\Theta$ being the x-ray wavelength and full scattering angle, respectively. The x-ray signal has rotational symmetry with respect to the incoming x-ray beam due to the averaging over all orientations of the microparticles. In standard processing of background subtraction, space angle correction, a one-dimensional scattering distribution I(q) is derived. In practice, two sets of data are recorded, one with the x-ray pulses arriving 2 nanoseconds before the laser pulses ("negative delay") and one with the x-ray pulses following the laser pulses by a defined delay to probe the time dependence of the dynamics. A difference $ \Delta$I(q) of the two signals contains only a signal pertaining to laser-induced structural changes in the sample. The observables deduced from this signal have been discussed in detail earlier \cite{ziefuss20,plech24nano}. In brief, the gold (111) powder peak can be analyzed in position, amplitude, and width to conclude on particle heating, melting and strain formation. Liquid scattering can provide information on the thermodynamic state of water. The full x-ray scattering intensity S(q) in a q range between 0.2 and 4.3 \rezAs is shown in fig. \ref{SEM}, b) together with the contribution from gold powder scattering, as well as a difference scattering curve at a delay of 45 ps and fluence of 2700 \Jm. Note the different amplitudes of the full scattering and the powder scattering of gold, the latter amounting only to 7 \% of the water background. Small-angle scattering finally reflects all changes in large-scale structure from few nm to about 60 nm in the present case. Previously, we have derived changes of particle sizes and shapes of irradiated initial nanoparticles \cite{plech17,plech24nano} as well as the quantification of the formation of vapor bubbles by cavitation \cite{kotaidis05apl,kotaidis06jcp}. 

However, compared to the q-values that indicate shape changes (Guinier range, q$_{Guinier}$ $\simeq \pi/r$ with r being the characteristic length scale as the radius), the microparticles scatter at q-values that were below the accessible detection range of the standard pinhole detector setup used in this study.  Small-angle scattering for compact particles generally decays with a power law of S(q) $\propto$ K q$^{-4}$ for q larger than the Guinier range independent of the actual r values and can therefore be detected in the accessible q window. The proportionality K can be used to express the total specific surface area A \cite{spalla03,schlumberger22}:

\begin{equation}
A = \frac{K}{2 \pi r_e^2 (N_A \cdot Z/(M \cdot \Delta \rho_{p-l}))}
\label{porod}
\end{equation} 

with the scattering cross section expressed by the classical electron radius $r_c$, Avogadro's constant $N_A$, atomic number Z, molecular mass M and electron density contrast $\Delta \rho_{p-l}$ between particle p and liquid l. Thus, the amplitude of the SAXS signal is proportional to the total surface area of the particles. The derivation of the amplitude is conveniently performed by averaging the experimental SAXS signal over a finite q range q$_P$ that obeys the Porod law. We chose the interval between 0.014 and 0.056 \rezAs to extract $\Delta$I(q$_P$). 

\begin{figure}[ht]
\centering
  \includegraphics[width=16cm]{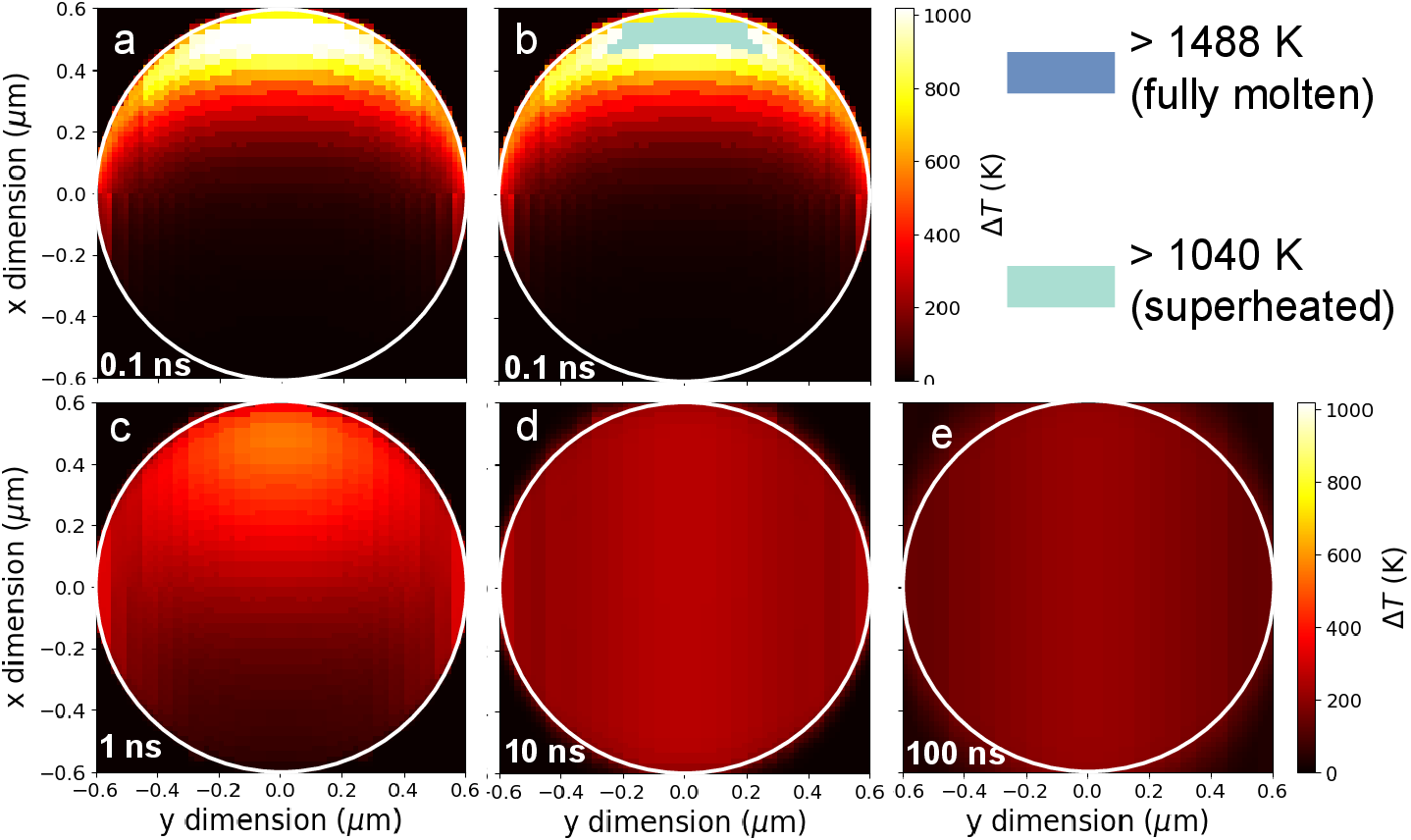}
  \caption{Numerical simulation (cross section through the center) of the heat transfer within a single gold microparticle upon irradiation from the top at a simulated fluence of 85 \Jm. The temperature is color coded, reaching 1170 K rise within 100 ps in a top layer facing the laser footprint (a), with full melting assumed, if the temperature exceeds a rise of 1488 K (see text) and in a superheated/partially molten state, if the temperature exceeds a rise of 1040 K to the melting point of bulk gold (b). The panels c) - e) show the temperature flow through the particles and later cooling at a delay of 1 ns, 10 ns and 100 ns, respectively.  }
  \label{map85}
\end{figure}

{\bf Heat flow in microparticles} Laser excitation and heat flow through the microparticles is simulated by using a numerical code for evaluating laser penetration, coupling to electrons and phonons (two-temperature model, TTM) and diffusional transport of hot electrons and phonons through the particle. The Python code NTMPy \cite{alber21} is freely available and codes this transport in thin films using an optional tilted incidence angle for the laser. The fluence input value therefore assumes the incident laser fluence, not only the part absorbed in the particle.  Here, a sphere of 0.6 \mums radius has been divided into 12 equal annular regions of fixed incidence angle and thickness.  The latter was chosen with respect to the thickness of the sphere at a given radius from the rotation axis parallel to the propagation of the plane-wave laser light. This approach is limited by neglecting polarization effects and not allowing for transverse heat flow (perpendicular to the annual slices). On the other hand, the transport of both electrons and phonons is correctly described, which matters for gold with a particularly long electron-phonon time and important electron transport \cite{hohlfeld00, plech24njp}. The assembly of these independently simulated slices produces the spatiotemporal distribution of heat in a photo-excited gold sphere, as shown in fig. \ref{map85}. For visual support, these 12 slices have been interpolated to 36 raster lines in the y dimension.

Threshold temperature changes are defined in a quasistatic model \cite{takami99,pyatenko13}. In order to reach the melting point of gold (1336 K) from laboratory conditions a temperature rise of 1040 K is needed, which can be done by providing energy from a defined fluence value. In order to fully melt the particles the latent heat has to be provided in addition. With 29.09 kJ/mol of energy to reach the melting point and the latent heat of 12.55 kJ/mol \cite{pyatenko13} the fluence has to be increased by 43 \%. At complete melting we connect this to a temperature rise of 448 K above the melting point. Thus, at a temperature rise in the simulations of a temperature rise of 1488 K has to be obtained to achieve complete melting calorically.

\section{Results and discussion}

\subsection{Thermal kinetics}

If the excitation depth is smaller than particle size and heat diffusion proceeds on a slower time scale than electron-phonon coupling or melting/freezing kinetics, the heat distribution in a laser-excited microparticle is not trivial.

Fig. \ref{map85} shows a cross-sectional heat map through a microparticle at several time delays after laser excitation by a Gaussian pulse of 1 ps. At 100 ps (part a) the lattice heating is localized in a surface region of about 200 nm facing the laser direction (from top). Already, the thickness of this heat-affected zone amounts to more than the laser penetration depth of about 30 nm at 400 nm in the simulation \cite{plech24njp}. This is a result of the onset of phonon diffusion (lattice heat conduction), but also caused by transport of heat by the excited conduction electrons prior to coupling to the lattice. At the chosen laser fluence one finds that the temperature at the front part of the microparticle has already reached a 1170 K rise and surpassed an increase by 1040 K of the melting point, while it stays below the temperature of 1488 K. 
Consequently, this region is marked in b) as superheated, as melting of the whole region would require the additional latent heat to convert into a liquid, which is not provided in full. Local melting may still happen. 

At 1 ns delay (c) the heat has penetrated further into the particle, while the fully molten region has shrunken because of the diffusional transport. At 10 and 100 ns (d and e) heat has spread throughout the particle and eventually will enter the surrounding water, leading to cooling of the particle. 

\begin{figure}[ht]
\centering
  \includegraphics[width=12cm]{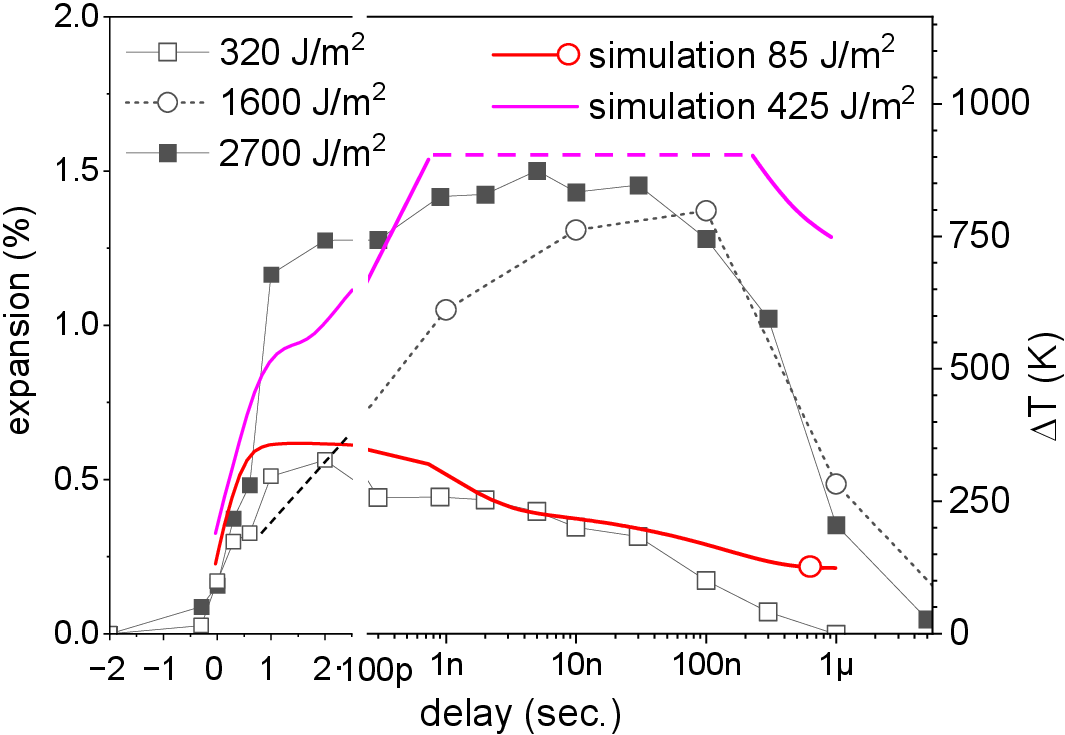}
  \caption{Time-resolved lattice expansion at several incident laser fluences of 320, 1600 and 2700 \Jms together with simulations of temperature rise $\Delta$T at corresponding fluences of 85 and 425 \Jms taking into account only parts of the volume showing a temperature increase of below 1040 K. At the simulated curve at 425 \Jms the volume fraction below a rise of 1488 K is too small for reliable averaging between 800 ps and 200 ns, thus marked by a dashed line at the maximum observable crystal temperature.}
  \label{kin}
\end{figure}

\begin{figure}[ht]
\centering
  \includegraphics[width=16cm]{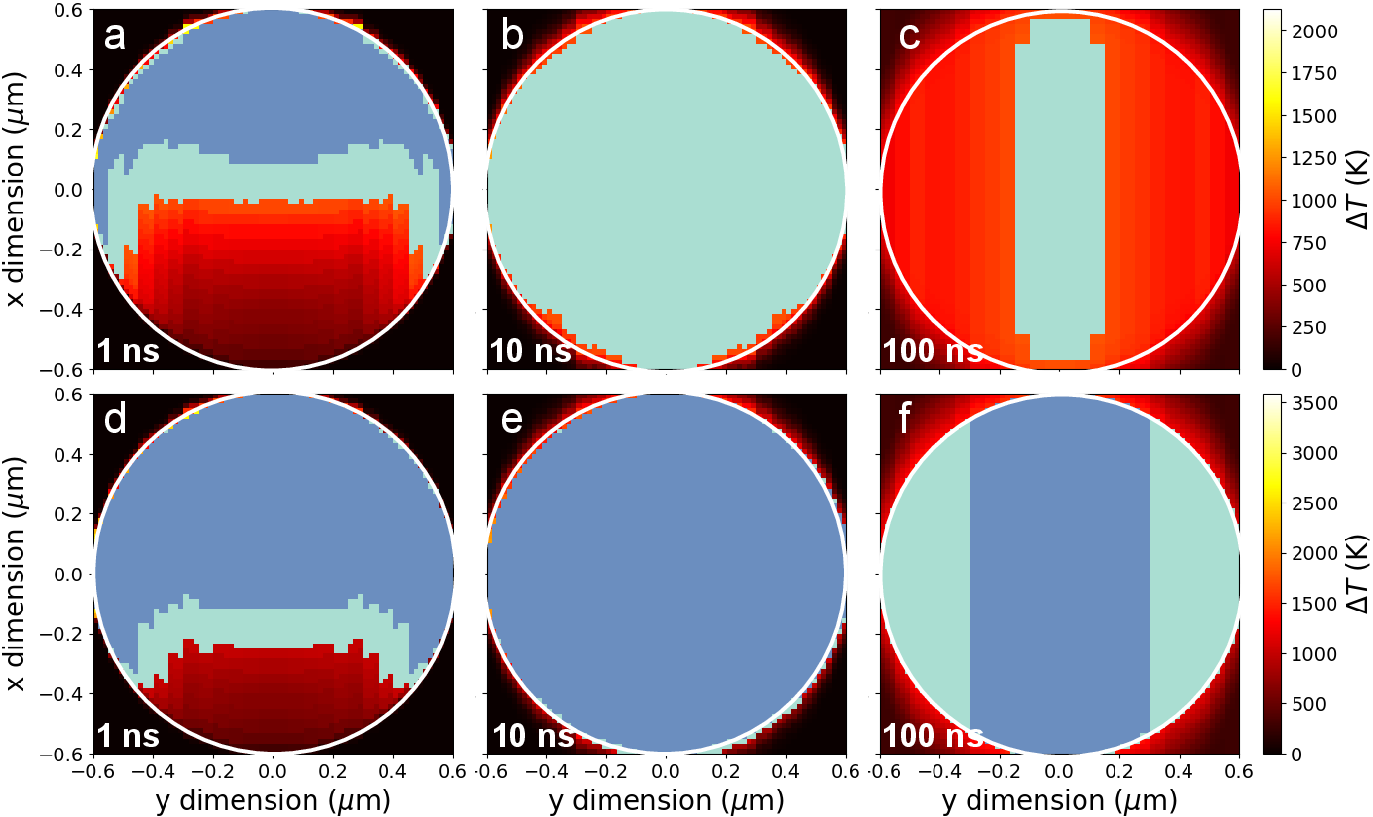}
  \caption{Numerical simulation of the melting state of a single gold microparticle as function of delay at a laser fluence of 425 (a - c) and 717 \Jms (d-f) with the color coding from fig. \ref{map85} marking fully molten regions in blue and superheated regions in light blue.}
  \label{map1600}
\end{figure}

Hereby, the simulated temperature from Fig. \ref{map85} were used to calculate an average particle expansion equivalent to the results of the diffraction experiment in Fig. \ref{kin}. For the  experiment, the particle expansion was determined from the average peak shift of the (111) powder peak at given fluence and delay and taking into account the lattice expansion coefficient $\alpha$ of 14.2$\cdot$10$^{-6}$/K at room temperature \cite{buffat76}.
The data is overlaid in figure \ref{kin} at a low experimental fluence of 320 \Jms, where particle melting is not yet observed, and higher fluences of 1600 and 2700 \Jms as function of delay represented by symbols. The average temperature at two simulated fluences of 85 and 425 \Jms (spatial temperature maps shown in figs. \ref{map85} and \ref{map1600} a-c) is overlaid as full lines. At 320 \Jms the expansion within the recorded time delay stays around 0.5 \%, which translates into a temperature change of 300 K or less. The temperature increases rapidly within 200 ps before reaching a plateau and decaying on a 100 ns time scale. This behavior is also reproduced by the simulation, which additionally allows the interpretation of front-side heating, spreading of the heat throughout the particles, and finally cooling to the water medium as compared to the volume-averaged experimental result. At 1600 \Jm, the observed experimental behavior differs substantially from the low-fluence excitation, as the temperature rise starts in the same way, but proceeds on a much longer delay to reach maximum temperature only between 10 and 100 ns.

A quantitative comparison shows that the simulation can reproduce the experimental behavior at 320 \Jms by a calculated fluence of 85 \Jms, which matches the change well in a delay region of 1-10 ns, where the heat is expected to have spread across the particle. The experimental and simulation fluence are off by a factor of 4. This is still satisfactory considering that the simulation model neglects polarization and also does not incorporate plasmonic effects that may redistribute energy along the particle surface or changes of optical properties with rising electron temperature \cite{plech24nano}. Both fluences are external fluences, i. e. quantifying  the fluence in front of the particle, but not the internal (absorbed) fluence. Additionally, in the experiment the reflection from the water surface of the jet is not taken into account, as well as extinction with depth in the jet, which lowers the fluence inside the jet. Also, phase transitions such as melting or fragmentation are not included, which consume energy. Indeed, the temperature distribution in the simulated maps in fig. \ref{map1600} shows that at 1 ns a large part of the front is molten and heated to much higher temperature.
The melting kinetics may have to be included, but it is generally acknowledged that melting at high excitation proceeds within tens of picoseconds, while melting could be delayed to 100-200 ps for homogeneous melting of single crystal materials \cite{rethfeld01,mo18}. On a longer time scale we would expect melting to instantly follow the temperature change. Between 1040 K and 1488 K rise we marked the respective area as superheated, which may include partial melting.
The energy spreads across the whole particles at later delays to partially or fully melt the particle at 100 ns. A small discrepancy between calculated absorptance and measured energy uptake has been seen earlier \cite{plech04prb,plech24nano,Spellauge2023} and could also be attributed to different loss mechanisms, such as nonlinear optical properties, thermal or acoustic coupling to the medium or laser shielding.

Contrarily, at 1600 \Jm, the lattice expansion apparently proceeds slower than at 320 \Jms and reaches maximum expansion only at 10 - 100 ns. The reason is not a modified thermal kinetics at this fluence, but an onset of melting on the picosecond time scale at the particle front. In this case, the molten volume fraction does not contribute to the powder scattering signal, but only the cold (and crystalline) back side. Therefore, the observed lattice temperature is much lower than the equivalent of dissipated energy in the microparticle.

After conduction of the thermal energy throughout the particle on a 10 ns time scale, the back side heats up as well.  Meanwhile, the front side starts to recrystallize and contribute again to the scattering signal. Therefore, the recorded temperature is highest on the nanosecond time scale before the particles cool due to heat release to water. This is also reflected in the simulation at 425 \Jm, where only the part of the particle below 1488 K was used for calculating the average temperature. The observed temperature is seen to stay at about 500 K below 0.5 ns to rise to higher values on the nanosecond scale. Between 800 ps and 200 ns most of the particle is molten such that the remaining solid part does not allow us to derive a meaningful average. Therefore this data is replaced by a dashed line. 

\begin{figure*}[ht]
\centering
  \includegraphics[width=16cm]{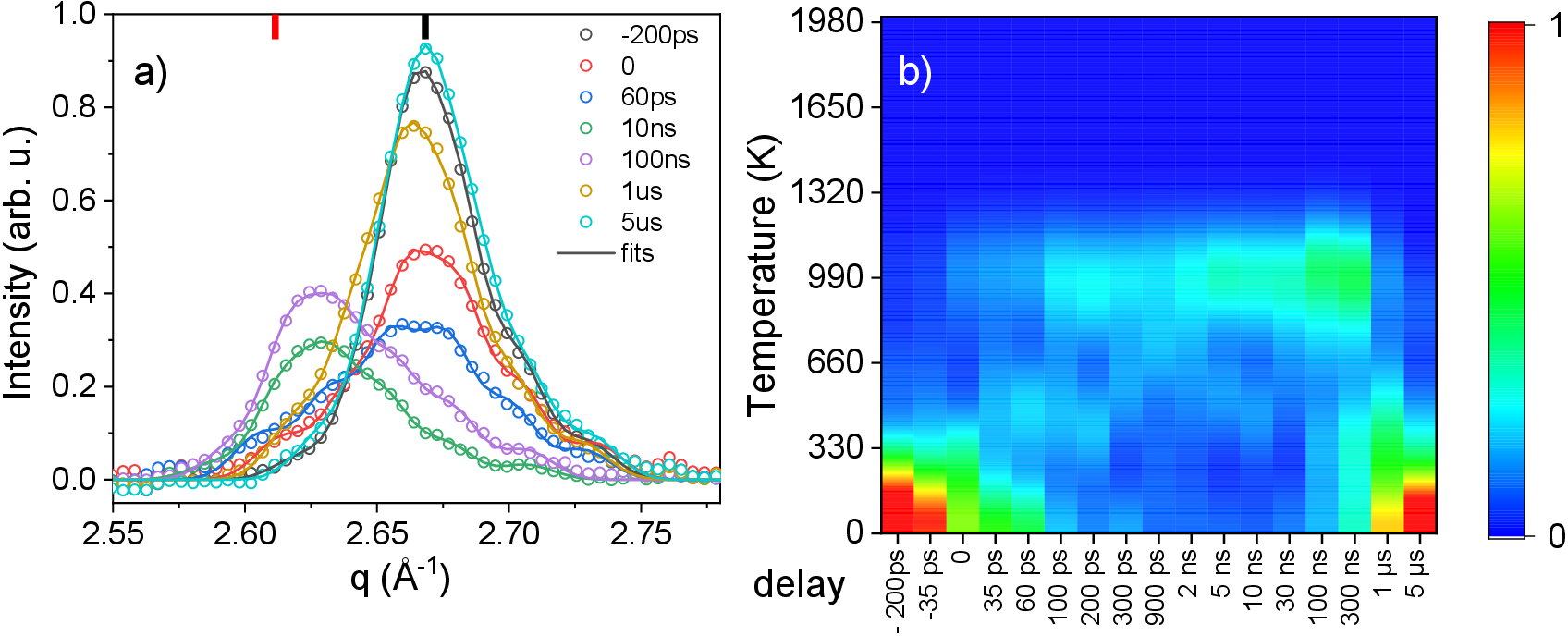}
  \caption{a) Powder profiles of the (111) reflection of gold as function of delay at 2700 \Jm. The lines are fits with a sum of Gaussian peaks. The vertical bars mark the position of the cold (111) peak (black) and the position of maximum expansion at the melting point. b) False color map of the extracted temperature-increase probability distribution as function of delay. The color scale maps the spectral weight of each temperature bin, see text for details. }
  \label{tempmap}
\end{figure*}

At 2700 \Jms the rise again is fast, but stays below to an expansion of 1.8 \%, the latter of which marks maximum expansion at the melting point. However, inspecting not only the peak shift, but also the peak intensity (fig. \ref{tempmap}), one recognizes that at 2700 \Jms the particles are molten to a large degree with the remaining crystalline volume staying close to the melting point. Therefore the measured temperature in the scattering experiment origins only from the parts of the particles that are still solid, why the expansion needs to stay below the melting expansion at 1.8\%.

In an alternative analysis approach, the (111) powder reflection can be viewed as a superposition of several narrow peaks that result from different parts inside the particle with varying temperature rise and thus a shifted peak position.  
The powder profiles extracted at an experimental fluence of 2700 \Jms are shown in fig. \ref{tempmap} a) for a sequence of time delays. While the non-excited peak (at delay - 200 ps) is found close to 2.675 \rezA, the peaks after excitation shift in position to lower q values, indicating lattice expansion. Additionally, the peak shapes change considerably because of possible melting (loss of peak intensity) and a spatially inhomogeneous temperature profile, as revealed by the simulations. For a quantitative analysis, the complete peak shape has been fitted by a superposition of six Gaussian peaks with fixed width but variable position and amplitude. The position of each is converted into temperature rise by using the thermal expansion coefficient (see above), and the amplitude contributes to the weight factor of each temperature. In order to reduce bias by the fit limits and starting values, we varied both for different fit runs (using Python code) and averaged over the results to convert the results into a histogram of temperature distribution within the particles. The implicit notion here is that a possible dispersion of excitation across the whole microparticle ensemble is neglected in favor of explaining the change of the powder profile in fig. \ref{tempmap} a) as a result of a distribution in temperature in a single particle.

The color map of the temperature-increase probability distribution as function of delay is shown in fig. \ref{tempmap} b. At negative delays the center of probability is located at zero temperature change as expected but shifts rapidly towards values between 0 and 1040 K with a broad distribution within.

Later, on a time scale of several ns the distribution narrows to a value close to the melting point at $\approx$ 1040 K. We interpret this observation in accordance to the simulation results where the temperature was also found to equilibrate throughout the whole microparticle (heat conduction) on the time scale of several nanoseconds. The spread in temperature on the picosecond time scale shows a large difference in temperature to the back part, which is reflected in an increased probability for a range of temperature changes. 
The temperature spread is subsequently reduced to show a more uniform distribution. Together with the fact that the powder intensity is reduced at the picosecond delays and recovers later, we conclude that the front part may have undergone partial melting and recrystallized when the heat spreads out. In this case, it is natural that, after release of latent heat, the temperature first equilibrates close to the melting point before the particle cools down via heat transfer to water. At the latest observed delay the temperature again settles at zero temperature change, indicating a recovery of the crystal lattice. One should add that the experiment is limited in resolving lattice defects or changes in crystalline particle size because of the limited resolution of the experimental setup.

\subsection{Particle fragmentation}

Thus, both the simulation and the time-resolved scattering show that a microparticle irradiated by a picosecond pulse is heated at the front side with a penetration depth of about 200 nm within the first 200 ps. The energy can cause the front side to melt and to dissipate the heat across the whole particle within 10 ns. The release of latent heat from the molten front side after recrystallization adds to the heating of the back side. Cooling to the water medium takes place with a time scale of 100 ns - 1 \mus and is potentially delayed at high fluence due to bubble formation and thermal shielding. The strong temperature gradient between front side and back side at delays of $<$ 1 ns leads to considerable stress that can cause thermomechanic fracture and particle fragmentation. However, the temperature at the front side can be high enough to cause phase explosion in gold when reaching the spinodal, which is the point of barrierless decomposition into atoms and clusters. With a typical spinodal temperature placed around 0.9 T$_c$, T$_c$ being the critical temperature of gold, we would expect spinodal decomposition at approximately an energy uptake of 4.4 eV/atom \cite{huang22,plech24nano}, which would be extrapolated to a necessary experimental fluence of 4700 \Jms to reach the spinodal temperature within the front layer of the microparticle. 

An additional fragmentation mechanism could be present, which concerns spallation of an excited front layer as been observed in laser ablation in liquids \cite{shih18,Spellauge2023,chen23}. This is described to take place close to the threshold for laser ablation in air \cite{sakurai21} or slightly higher in liquid, which for gold happens at about 22 k\Jms \cite{spellauge2022}. 

\begin{figure*}[ht]
\centering
  \includegraphics[width=17cm]{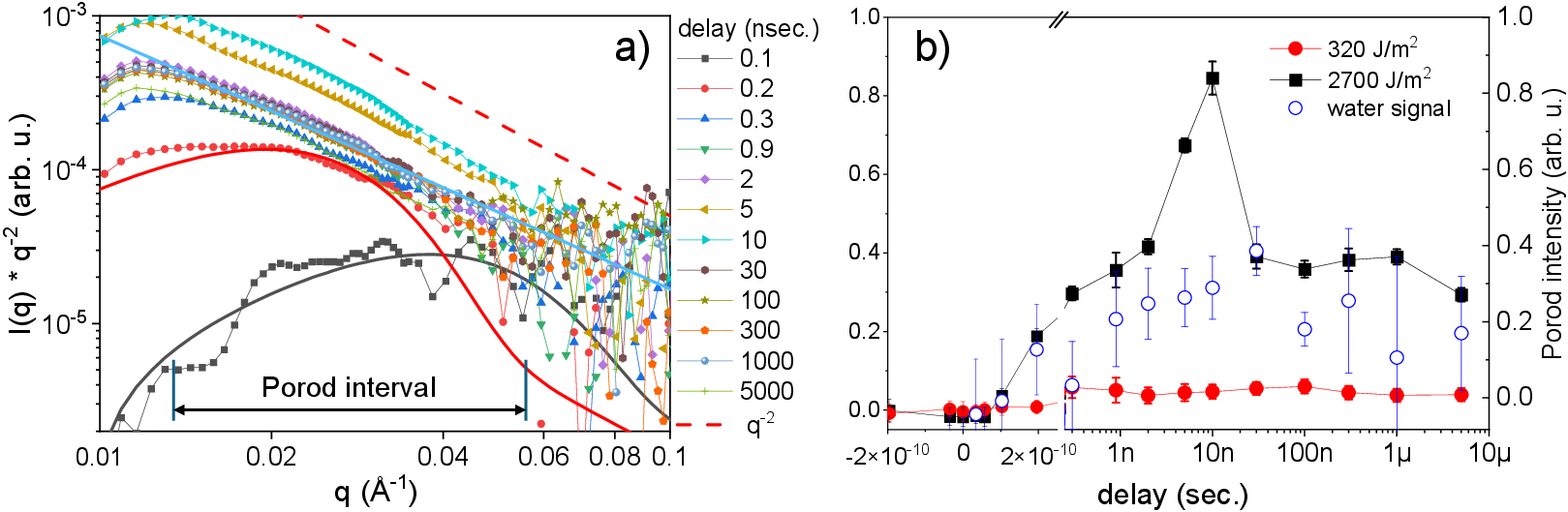}
  \caption{ a) Kratky representation of the small-angle part of the difference scattering after laser excitation at a fluence of 2700 \Jm. The dashed red line marks a slope of -2 for comparison. The solid lines are fits to the data at delay 0.1 ns (black), 0.2 ns (red) amd 1 \muss (pale blue) as explained in the text.  b) Change of the integrated (Porod) intensity as function of delay at 320 \Jms (red bullets) and 2700 \Jms (black squares) together with the amplitude of water difference scattering indicating bubble formation (blue circles) at 2700 \Jm. 
  %[Parameters diam. 12 nm, 1 \% at 0.1 ns, 23 nm 2.5 \% at 0.2 ns and 240 nm, 100 \% at 1 us]
  }
  \label{poroddelay}
\end{figure*}

Changes in scattering signal are detectable not only around the powder peak angles of crystalline gold, but also at other scattering vectors. Liquid scattering changes with characteristic functions upon heating and compression/expansion due to slight changes of interatomic distances of water \cite{cammarata06,kotaidis05apl,plech24nano}. The observed changes for microparticle excitation are weak (see fig. \ref{poroddelay}), but visible and point towards bubble formation around the heated particles and as reaction to that the compression of bulk water. More prominent changes are seen in the small-angle region, which corresponds to larger length scales and reflects particle shape change in general. In the q range from 0.01 to 0.1 \rezAs we observe a drastic increase in scattering as shown in fig. \ref{poroddelay}. At 2700 \Jms the scattering increases at low q as function of delay. For the earliest delays (100 ps, black, and 200 ps, red curves in fig. \ref{poroddelay}) the shape of the function $\Delta$I(q)$\cdot$q$^2$ shows a maximum around 0.035 resp. 0.02 \rezA, while for delays $>$ 200ps the shape is monotonic and changes in amplitude, but not in shape. The slope of this signal shows a power law decay of q$^{-2}$, which clearly indicates that $\Delta$I(q) scales with q$^{-4}$, indicative of the Porod region of compact, solid objects. Therefore, it is concluded that the microparticles are being fragmented to form new particles with increased surface area. As the shape of the curves at these later delays does not change from power law, it is not possible to derive a size distribution but one can assume that the formed fragments are still larger than the resolution of the present setup of about 60 nm. At the same time, the increase in surface area according to eq. \ref{porod} can be extracted as being proportional to the integrated intensity. 

\begin{figure}[ht]
\centering
  \includegraphics[width=12cm]{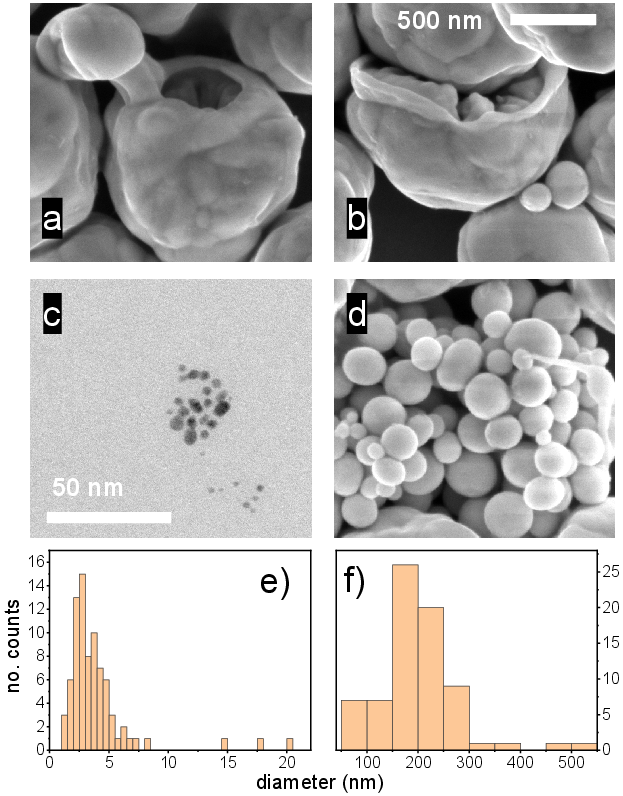}
  \caption{A set of SEM images of the fragmentation products at experimental fluence of 2700 \Jms (a,b,d) and a TEM image of the small fragments (c) together with the size histograms of two fraction of fragments from 0 -25 nm (e) and 75 -550 nm (f).}
  \label{SEMsize}
\end{figure}

The difference scattering signal at 100 ps and 200 ps has been fitted with fragments of different sizes. The experimental data are best matched when assuming an average particle radius of 12 nm or 23 nm in diameter, respectively. However, the nanoparticles only account for 1 \% and 2.5 \% of the total mass. This shows that small particles are emitted from a small volume fraction of the microparticles before the whole particle breaks up. This may be identified with extreme superheating at the front of the microparticles. At later delays ($\ge$ 300 ps) this signal is still present (as we can observe clusters in the range of 2-20 nm in TEM (see fig. \ref{SEMsize}), but hidden below the larger signal from the large fragments. A simulation at a fluence comparable to the experimental fluence of 2700 \Jms results in temperature rise at the irradiated front of the microparticles of 9700 K, which approaches the spinodal locally \cite{boborides99}. At these later delays, the signal from these first fragments is superseded by larger fragments as described above.

The amplitude of the SAXS signal $>$ 200 ps is converted into an increase in surface area, when an initial microparticle of radius R$_0$ is fragmented into N smaller fragments of radius R$_{frag}$. An estimate of the size of the fragments can be given under the assumption of spherical fragments, assuming that the total mass V$_{total}$ did not change:

\begin{equation}
V_{total} = 4 \pi / 3 \cdot R_0^3 = N \cdot 4 \pi / 3 \cdot R_{frag}^3
\label{eqvol}
\end{equation}

Then the increase in surface area S$_{frag}$/S$_0$ can be expressed as:

\begin{eqnarray}
S_{frag}/S_0 & = & N \cdot R_{frag}^2 / R_{0}^2 \nonumber \\ 
& = & R_0 / R_{frag}, \qquad \mbox{(using eq. \ref{eqvol})} \\
\mbox{and:}&  & \nonumber \\ 
S_{frag}/S_0 & = & (\Delta I(q_P) +I_0(q_P)) / I_0(q_P) \label{surf}
\end{eqnarray} 

by using the amplitude of the SAXS signal $\Delta$ I(q$_P$) and the amplitude of the scattering from the initial microparticles I$_0$. The latter is derived from a normalization of  $\Delta$ I by a colloid of known size and concentration (see \cite{plech24nano}) rather than using the raw scattering.

Following this approach, we estimate that the initial surface of the microparticles has increased by a factor of 10 upon fragmentation at a delay of 1 \mus. This would result in a diameter of 80 nm of presumed spherical products. However, thermomechanic fragmentation of solid particles can lead to nonspherical shapes, which show a higher surface-to-volume ratio, and thus can be larger in size. At two delays of 5 and 10 ns the SAXS intensity is further increased beyond the scattering at 1 \mus, which we consider the final state of fragmentation, because of the coinciding cooling time. The SAXS signal amplitudes are displayed in fig. \ref{poroddelay} b) at 2700 \Jms and the lower value of 320 \Jm, at which the particles stay in the solid state. At 320 \Jms barely any difference signal is observed, which excludes fragmentation at this fluence. Additionally, the amplitude of the water difference signal is shown (open, blue circles in fig. \ref{poroddelay} b) at 2700 \Jm, which starts to rise at 200 ps and does not fully decay within the observed delay range. As we relate this signal to bubble formation, it shows that vapor bubbles around the hot particles are formed and long-lived. Spellauge et al. \cite{Spellauge2023} record bubbles around individual irradiated IrO$_2$ particles that have a life time of $>$ 10 \mus, which agrees well with our findings. 

The SAXS signal in fig. \ref{poroddelay} b) shows a rise time $>$ 200 ps, which may reflect the finite disintegration speed of a microparticle into spatially separated fragments. The SAXS signal reaches a constant value on the nanosecond scale, indicating a stationary state. This is only interrupted by a signal maximum at 5 and 10 ns as mentioned above. We believe that the nascent bubbles around the particles contribute to the signal at these time delays strongly if their size is only slightly larger than the resolution limit, thus causing a relatively sharp peak in delay. 

\begin{figure}[ht]
\centering
  \includegraphics[width=12cm]{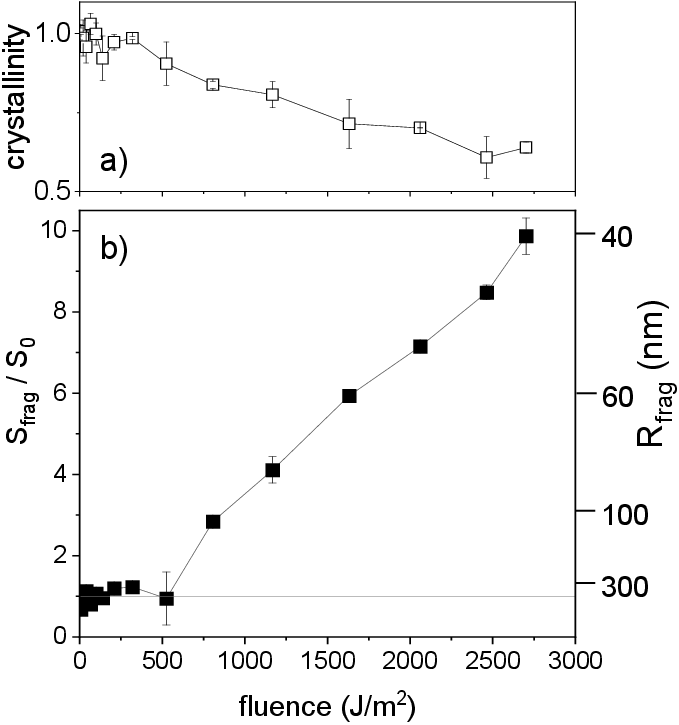}
  \caption{a) Solid fraction of microparticle mass at 1 - 10 ns delay as function of fluence. b) Increase in surface area after fragmentation derived from  the amplitude of the excess small-angle scattering within the Porod interval as marked in fig. \ref{poroddelay} as function of fluence. The right axis in b) indicates a corresponding particle radius, if spherical fragments are assumed.}
  \label{porodfluence}
\end{figure}

The more interesting part is the steady-state scattering at 1 \muss that represents the final fragment structure after particle cooling. At this fixed delay, the fluence was increased stepwise to follow the possible onset of fragmentation. The results are displayed in figure \ref{porodfluence}. As discussed above, the difference scattering curves show a constant slope in the SAXS region that does not allow to apply shape-dependent models for the simulation. The increase in surface area was derived according to eq. 4. %\ref{surf}. 
The surface area increases in total by a factor of 10 from the initial microparticle surface to that of a number N of fragments, as seen above in intensity change of the time-resolved data. Thus, fragments of about 80 nm diameter can be formed, which is equivalent to N = 3400 fragments formed. Again, we have no information about the shape, which renders N to an approximate number. 

The inspection of the SEM and TEM images of the collected colloid reveals partly fragmented microparticles and fragments of variable size, see fig. \ref{SEMsize} a-d. A number of the fragments shows a large variation in diameter from 75 to 550 nm as seen in part d and the histogram in part f. Some small fragments can also be found after TEM inspection with particle diameters in the 5 nm range (see TEM image in c and histogram in e). No statement on relative frequency or fragmentation efficiency can be made from the electron microscopy images, but it is intriguing to find repeated damaged microparticles of apparently initial diameter but an eroded side with a deep indentation. This supports the notion that strain at the irradiated front of the particles plays a major role in fragmentation, whereas the back side seems to remain at low temperature and does not show damage. The large fragments, on the other hand, show a compact round shape, which is indicative of at least partial melting. This indicates that the fragmentation mechanism is thermoelastic with thermally induced strain leading to eruption of the front part and emission of partially molten material that forms particles in the 100-500 nm range, which agrees with the SAXS results. Some small fragments in the 5 nm diameter range may be formed at the front surface, where the temperature reached a multiple of the melting temperature. Similar observations have been reported by Spellauge et al. \cite{spellaugeunpublished} based on single-particle ultrafast imaging. They report the dominance of photomechanical fragmentation at low fluence and onset of photothermal (phase explosion) fragmentation at 15 times the photomechanical fragmentation threshold.  This mechanism thus shows similarities to laser ablation in liquid, with the heat-affected zone being much smaller than the irradiated target.

The fluence dependence shows that below 500 \Jms no detectable surface increase is found, which puts the experimental fragmentation threshold at $<$ 750 \Jms. At this fluence, we expect the surface temperature on the sub-nanosecond time scale to reach 2600 K at maximum, which raises a volume fraction of 5-10 \% of the microparticle above melting for some time after laser excitation, while the back part of the particle does not differ much from room temperature. At 100 ns, the maximum temperature rise has decayed to 500 K throughout the microparticle. This infers that substantial strain prevails within the microparticle for a delay of up to 1 ns. Note that the acoustic relaxation time $\tau_{acoust}$ across the particle diameter D = 1.2 \mums  is D/v$_s$ with v$_s$ being the speed of sound of gold \cite{lide95}. Thus, the relaxation time for compression waves is 370 ps, for shear waves 1 ns, which is in good agreement with the previous interpretation. This shows that at least on 300 ps time scale, stress is confined in the particle and can cause fragmentation. The crystalline fraction averaged over 1 -10 ns as displayed in fig. \ref{porodfluence}  a) reflects the fact that melting occurs to a variable fraction at the front side, but does not happen across the whole particle. At the highest observed fluence only about 50 \% of the particles mass is molten, which could in part also come from an inhomogeneous fluence profile throughout the liquid volume.  

From the fragmentation threshold at $<$ 750 \Jms the surface area increases almost linearly, while the fragment size decreases in an accelerated fashion with fluence.

\section{Conclusions}

The fragmentation of microparticles by pulsed lasers has become an alternative route to producing defined nanoparticles from a commercially available precursor. It is difficult to resolve the mechanisms of fragmentation by experimental approaches, which leaves conflicting reports on this process in the literature. We have used 1 ps laser pulses at 400 nm to photo-excite a colloid of 1.2 \mums gold microparticles and watch fragmentation by atomically resolving methods of x-ray scattering with 60 ps time resolution. The powder scattering resolved the dynamics of the temperature of the microparticles, either as average over the whole particle or even resolved as a spectrum of temperature by fine-analyzing the powder scattering profiles. We observe the conventional phenomenon of heating the particles on a 100 ps time scale and subsequent cooling on a 100 ns - 1 \muss time scale at low applied laser fluence, but an unconventional behavior at higher fluence, where the heating seems to be delayed by 1-5 ns with respect to laser impact. This behavior can be explained aided by simulational calculation of the spatio-temporal distribution of heat across a spherical particle by full solution of the TTM and heat diffusion in this geometry. These simulations reveal that for the first 100-200 ps the particle is only heated at the front side with a penetration depth of 200 nm. This depth results from a combination of optical penetration depth and diffusional energy transport. Despite an electron-phonon coupling of 5-10 ps \cite{guzelturk20}, which favors phonon transport at times beyond this coupling time, electrons may still move ballistically or diffuse some 100 nm before equilibrating with the phonon bath \cite{rethfeld01,plech24njp}. The average temperature of the particles rises by about 250 K for an applied fluence of 320 \Jm. 

\begin{figure*}[h]
\centering
  \includegraphics[width=15cm]{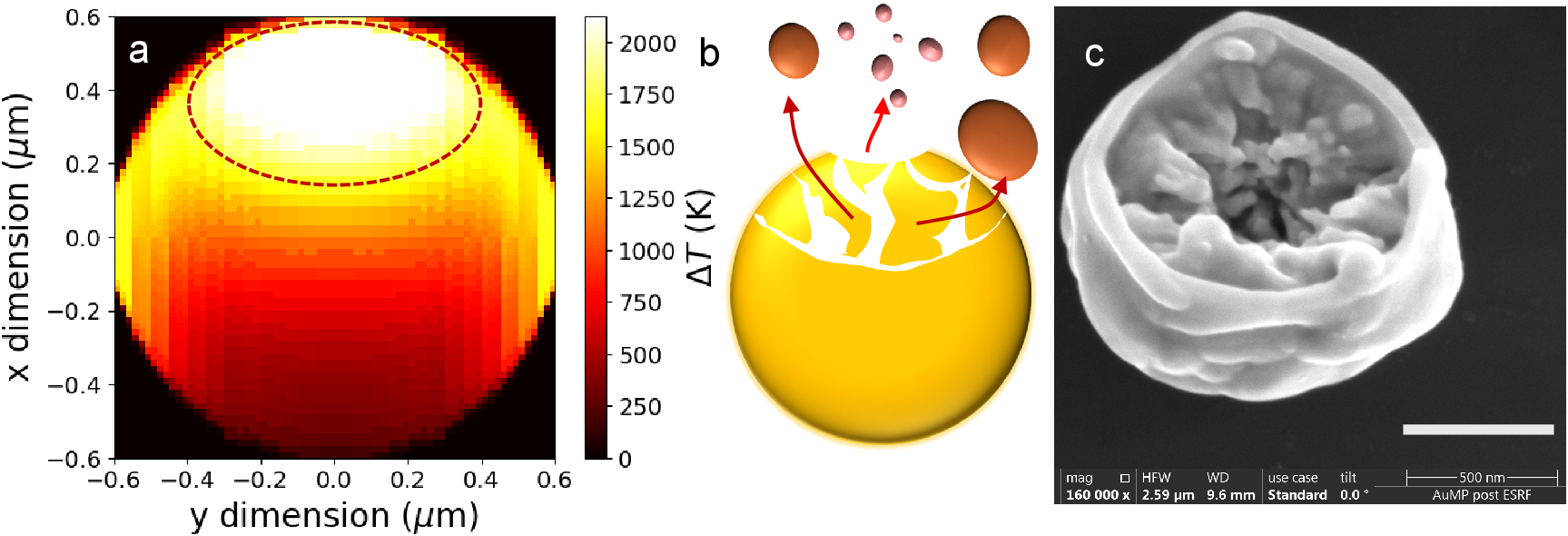}
  \caption{a) Temperature distribution at 1 ns of the simulated microparticle excitation at 717 \Jms with the region of the highest temperature marked by a dashed line. b) Sketch of the proposed mechanism with thermoelastic fracture of the part the region of highest temperature together with the generation of few-nm particles at the very front side after phase explosion. c) Example of an SEM image of an irradiated microparticle with a central cavity that may be linked to the thermoelastic fracture. }
  \label{mech}
\end{figure*}

At high enough fluence (for example, 1600 \Jm) the temperature at the front side of the irradiated microparticles is high enough to melt a part of the particle mass, which then would not contribute to powder scattering any more and is thus not included in the temperature evaluation. Only after heat diffusion and recrystallization will this energy be transported to the back side and detected as lattice expansion. 

Fragmentation follows a thermoelastic driving mechanism, which is illustrated in fig. \ref{mech}. The SAXS region of the scattering from excited particles shows distinct signals that include small-angle scattering from nanoparticles in the 12-24 nm range at a delay of 100 -200 ps, probably emitted from the front side during excess heating and melting of this surface part. At later delays and high fluence the SAXS signal is stationary with a slope of q$-4$, but changing amplitude. This indicates that new particles are formed that exceed the resolution limit of the experiment of 60 nm to be resolved in size. However, additional scattering is proportional to the change in surface area. Therefore, a progressing fragmentation of the microparticles can be described that increases with fluence and delay. Beyond about 1 ns the scattering stays stationary, indicating the completion of the fragmentation process. The signal amplitude can be used to estimate the fragment size, which is found to be about 80 nm, if spherical fragments are assumed. We find a fragmentation threshold of $<$ 750 \Jm, at which a first sign of particle rupture is found to end in few pieces, which matches well the 370 \Jms of absorbed fluence as found by Spellauge et al. \cite{spellaugeunpublished}. Up to 2700 \Jms the number of fragments increases to 3400, making this process very effective. The number of very small nanoparticles is still low (1 - 2.5 \%), while the produced nanoclusters are probably below the detection limit due to the dominance of the signal by the large fragmented particles. To increase the yield of these clusters, a higher fluence has to be applied, which will drive the overall particle temperature closer to evaporation and spinodal decomposition to cause phase explosion. The present thermal kinetics, the time scale of the fragmentation, and fragment sizes point towards a mixture of two mechanisms, the  stress-induced fragmentation, where parts of the microparticles may still be in the solid state, and the forces arising from the temperature gradient between front and back sides complemented by photothermal fragmentation at much higher fluence than the threshold for photomechanical fragmentation.

\section*{Author contributions}

The research program was conceived and designed by AP and SR. The samples have been prepared by MT. The experiment has been conducted by all authors. Analysis was performed by YP and AP. The manuscript has been initially written by YP and AP with contributions from all authors. 

\section*{Conflicts of interest}
There are no conflicts to declare.

\section*{Data availability}

The raw data are available after the expiration of the embargo (2027) at https://doi.org/10.15151/ESRF-ES-1560212245. Additionally, this data and reduction routines (Python scripts) are available upon request from anton.plech@kit.edu.

\section*{Acknowledgements}

This work is supported by the German Science Foundation DFG under contract 491072288 and by the Helmholtz programme "From Matter to Materials and life". We acknowledge the European Synchrotron Radiation Facility (ESRF) for provision of synchrotron radiation facilities under proposal number SC5562. We would like to thank Maximilian Spellauge for useful discussions.

%%%END OF MAIN TEXT%%%


%\balance

\renewcommand\refname{References}

%%%REFERENCES%%%
\bibliography{rsc} 
\bibliographystyle{MSP}

% Table of contents entry should be 50 - 60 words long
% Image should be 55 mm broad and 50 mm high or 110 mm broad and 20 mm high

\begin{figure}
\textbf{Table of Contents}\\
\medskip
  \includegraphics[width=11cm]{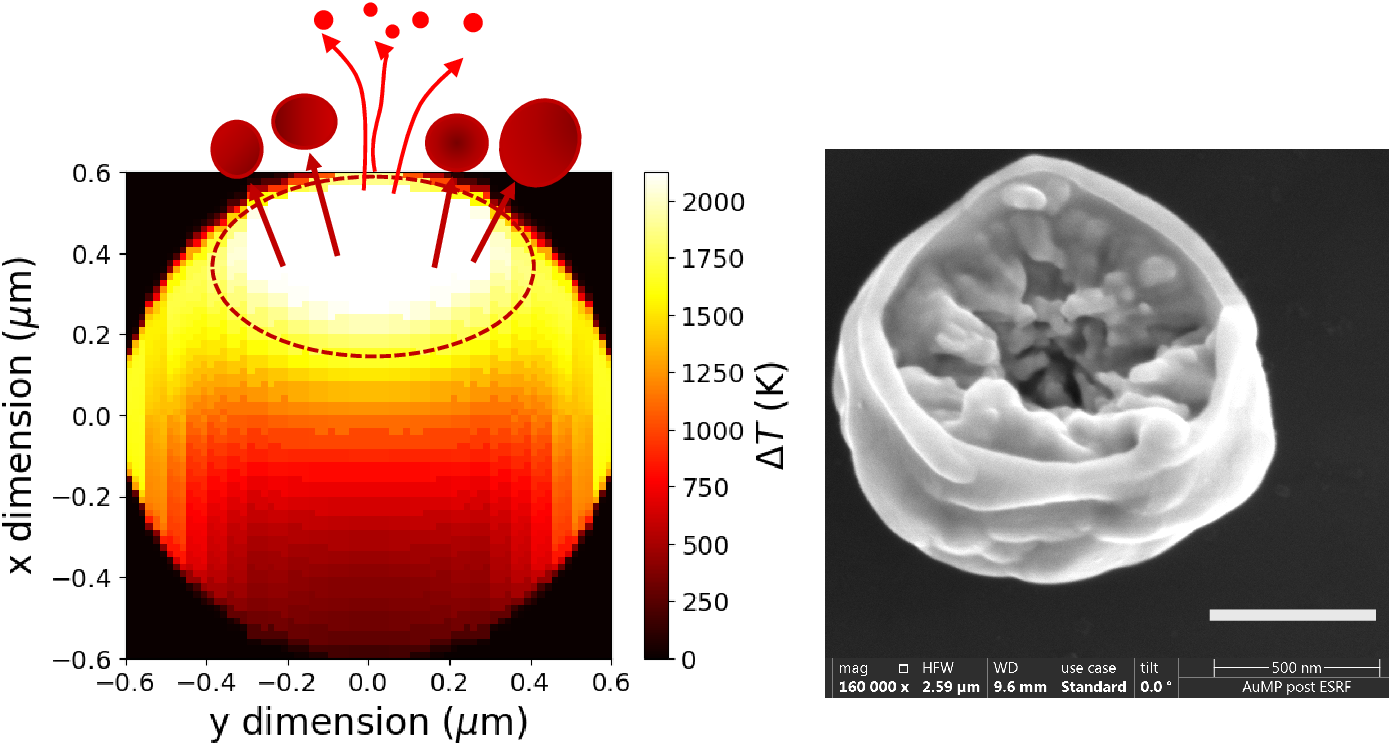}
  \medskip
  \caption*{Gold microparticles in liquid are irradiated by picosecond laser pulses to induce melting and fragmentation. The time-resolved x-ray scattering results show a photomechanic fragmentation threshold followed by photothermal fragmentation at higher fluence.}
\end{figure}

\end{document}